\documentclass[12pt,letterpaper]{article}
\usepackage[top=0.85in,left=2in,footskip=0.75in,marginparwidth=2in]{geometry}

\usepackage[utf8]{inputenc}
\usepackage{cite}
\usepackage{nameref}
\usepackage[right]{lineno}
\usepackage{amsmath}
\usepackage{microtype}
\DisableLigatures[f]{encoding = *, family = * }

\raggedright
\usepackage{changepage}

\usepackage[aboveskip=1pt,labelfont=bf,labelsep=period,singlelinecheck=off]{caption}

\makeatletter
\renewcommand{\@biblabel}[1]{\quad#1.}
\makeatother

\usepackage{lastpage,fancyhdr,graphicx}
\usepackage{epstopdf}
\usepackage{color}

\definecolor{Gray}{gray}{.25}

\usepackage{graphicx}

\usepackage{sidecap}

\usepackage{wrapfig}
\usepackage[pscoord]{eso-pic}
\usepackage[fulladjust]{marginnote}
\reversemarginpar

\begin{document}
\vspace*{0.35in}

% title goes here:
\begin{flushleft}
{\Large
\textbf\newline{Time-of-Flight Three Dimensional Neutron Diffraction in Transmission Mode for Mapping Crystal Grain Structures}
}
\newline
% authors go here:
\\
Alberto Cereser \textsuperscript{1,2,*}
Markus Strobl \textsuperscript{2,3}
Stephen Hall \textsuperscript{4,2}
Axel Steuwer \textsuperscript{5,6}
Ryoji Kiyanagi \textsuperscript{7}
Anton Tremsin \textsuperscript{8}
Erik Bergb\"ack Knudsen \textsuperscript{1}
Takenao Shinohara \textsuperscript{7}
Peter Willendrup \textsuperscript{1}
Alice Bastos da Silva Fanta \textsuperscript{9}
Srinivasan Iyengar \textsuperscript{10,2}
Peter Mahler Larsen \textsuperscript{1}
Takayasu Hanashima \textsuperscript{11}
Taketo Moyoshi \textsuperscript{11}
Peter M. Kadletz \textsuperscript{12}
Philip Kroo{\ss} \textsuperscript{13}
Thomas Niendorf \textsuperscript{13}
Morten Sales \textsuperscript{1}
Wolfgang W. Schmahl \textsuperscript{12}
S{\o}ren Schmidt \textsuperscript{1,$\dagger$}
\\
\bigskip
\bf{1} {\footnotesize{NEXMAP}}, Department of Physics, Technical University of Denmark, Kgs. Lyngby, 2800, Denmark
\\
\bf{2} European Spallation Source {\footnotesize{ESS AB}}, Lund, 22592, Sweden
\\
\bf{3} Niels Bohr Institute, University of Copenhagen, Copenhagen, 2100, Denmark
\\
\bf{4} Division of Solid Mechanics, Lund University, Lund, 22362, Sweden
\\
\bf{5} Nelson Mandela Metropolitan University, Port Elizabeth, 6031, South Africa
\\
\bf{6} University of Malta, Msida, {\footnotesize{MSD}} 2080, Malta
\\
\bf{7} {\footnotesize{J-PARC}} center, Japan Atomic Energy Agency, Tokai-mura, 319-1195, Japan
\\
\bf{8} Space Sciences Laboratory, University of California at Berkeley, Berkeley, California 94720, {\footnotesize{USA}}
\\
\bf{9} Center for Electron Nanoscopy, Technical University of Denmark, Kgs. Lyngby, 2800, Denmark
\\
\bf{10} Division of Materials Engineering, Lund University, Lund, 22362, Sweden
\\
\bf{11} Research Center for Neutron Science and Technology, {\footnotesize{CROSS}}, Tokai, Naka-gun 319-1106, Japan
\\
\bf{12} Applied Crystallography and Materials Science, Department of Earth and Environmental Sciences, Ludwig-Maximilians-Universit\"at, M\"unchen, 80333, Germany
\\
\bf{13} Institut f\"ur Werkstofftechnik (Materials Engineering), Universit\"at Kassel, Kassel, 34125, Germany
\\
\bigskip
* alcer@fysik.dtu.dk
\\
$\dagger$ ssch@fysik.dtu.dk

\end{flushleft}

\section*{Abstract}
The physical properties of polycrystalline materials depend on their microstructure, which is the nano-to-centimeter-scale arrangement of phases and defects in their interior. Such microstructure depends on the shape, crystallographic phase and orientation, and interfacing of the grains constituting the material. This article presents a new non-destructive {\footnotesize{3D}} technique to study bulk samples with sizes in the cm range with a resolution of hundred micrometers: time-of-flight three-dimensional neutron diffraction ({\footnotesize{ToF 3DND}}). Compared to existing analogous {\footnotesize{X}}-ray diffraction techniques, {\footnotesize{ToF 3DND}} enables studies of samples that can be both larger in size and made of heavier elements. Moreover, {\footnotesize{ToF 3DND}} facilitates the use of complicated sample environments. The basic {\footnotesize{ToF 3DND}} setup, utilizing an imaging detector with high spatial and temporal resolution, can easily be implemented at a time-of-flight neutron beamline. The technique was developed and tested with data collected at the Materials and Life Science Experimental Facility of the Japan Proton Accelerator Complex ({\footnotesize{J-PARC}}) for an  iron sample. We successfully reconstructed the shape of 108 grains and developed an indexing procedure. The reconstruction algorithms have been validated by reconstructing two stacked Co-Ni-Ga single crystals, and by comparison with a grain map obtained by post-mortem electron backscatter diffraction ({\footnotesize{EBSD}}). 

% now start line numbers
%\linenumbers

% the * after section prevents numbering
\section*{Introduction}

Polycrystalline materials, abundant in nature and among man-made objects, are aggregates of grains joined by a network of internal interfaces. The macroscopic properties of these materials are mostly defined by their microstructure and by micro-structural processes. Consequently, to understand the behavior of such a material, it is crucial to probe its internal structures, which range over a number of length scales \cite{clemens2008microstructure}.

Standard tools in metallography, such as optical and electron micrography, return information limited to the microstructure of a sample surface, which may not be representative of the bulk material \cite{surf_meas, haynes2013optical}. These techniques require extensive sample preparation and can return {\footnotesize{3D}} sample reconstructions only by repeatedly removing a layer of material and characterizing the surface beneath \cite{alkemper_2001, zankel_2014}. Such processes are destructive and cannot be applied {\emph{in situ}}, e.g. to map structural changes during loading.

In the last twenty years, various nondestructive techniques have emerged to study {\footnotesize{3D}} shape and orientation of the grains composing polycrystalline materials. At first, techniques employing {\footnotesize{X}}-rays were developed: three-dimensional {\footnotesize{X}}-ray diffraction microscopy ({\footnotesize{3DXRD}}), diffraction contrast tomography ({\footnotesize{DCT}}) and high energy {\footnotesize{X}}-ray diffraction ({\footnotesize{HEDM}}) \cite{poulsen2001three, DCT1, suter2006forward}. {\footnotesize{3DXRD}}, {\footnotesize{DCT}} and {\footnotesize{HEDM}} allow investigation of micrometer- to millimeter-sized samples with resolutions ranging from tens of nanometers to micrometers. 

More recently, neutron diffraction contrast tomography (n{\footnotesize{DCT}}) has been developed to study, at continuous neutron sources, millimeter-scale samples {\cite{peetermans2014cold}}. In n{\footnotesize{DCT}}, the sample is illuminated with a continuous, polychromatic neutron beam, and the diffracted signal is collected in backscattering Laue mode. The number and size of the grains that can be reconstructed using n{\footnotesize{DCT}} is limited by diffraction spots overlapping and blurring, which set a minimum grain size of 1 mm for a mosaicity of 0.1-0.2$^{\circ}$ \cite{peetermans2014cold}. Despite providing a lower resolution than their X-ray counterparts, neutron imaging techniques have the advantage of being able to probe the bulk of millimeter-to centimeter sized samples. Moreover, the higher penetration capabilities of neutrons compared to {\footnotesize{X}}-rays are particularly important for engineering materials. 

Here we present time-of-flight three-dimensional neutron diffraction ({\footnotesize{ToF 3DND}}). Compared with {\footnotesize{X}}-ray techniques, {\footnotesize{ToF 3DND}} allows to study larger samples and, through its ability to profit from pulsed spallation sources, has the potential to improve the resolution for neutron techniques in both space and in time. {\footnotesize{ToF 3DND}} utilizes a conventional imaging geometry in time-of-flight mode, providing intrinsic neutron energy resolution that enables the reconstruction of the {\footnotesize{3D}} shape and orientation of the grains composing polycrystalline materials. The approach recalls the direct-beam {\footnotesize{X}}-ray {\footnotesize{DCT}} technique developed by Ludwig {\emph{et al.}} \cite{DCT1}. {\footnotesize{ToF 3DND}} can be implemented both at dedicated {\footnotesize{ToF}} imaging beamlines and at {\footnotesize{ToF}} neutron diffractometers. In the latter case, the simultaneously available diffraction data may add significant information about individual grain strain states, mosaicity or twinning.

\section*{Principle}

The attenuation of a neutron beam through a sample is the mechanism enabling transmission imaging. The attenuation by a specific material can be described by the {\emph{linear attenuation coefficient}} $\Sigma$, which depends on the density $N$ of the involved nuclei and on their absorption and scattering cross sections, $\sigma_a$ and $\sigma_s$ \cite{fermi1947transmission}

\begin{equation}
\Sigma = N(\sigma_a + \sigma_s)
\end{equation}

The total scattering cross section $\sigma_s$ is a sum of the different elastic and the inelastic contributions, both consisting of a coherent ($\sigma_{ela}$) and incoherent ($\sigma_{inela}$) part. In the range of cold and thermal neutrons, generally used for neutron imaging and diffraction since their wavelengths match the crystal lattice distance, for many crystalline materials the attenuation coefficient is dominated by the coherent elastic scattering cross section.

For a single bound nucleus, the total coherent scattering cross section can be written as $\sigma_{coh} = 4\pi b_n^2$, with $b_n$ being the bound coherent scattering length. For crystals, which consist of a matrix of specifically ordered nuclei, the significant part of the cross section takes the form

\begin{equation}
\sigma_{coh, ela}(\tau) = \frac{4\pi^3 N}{k^2 V_0}\sum_{\tau}^{\tau < 2k}\frac{1}{\tau} |F(\tau)|^2
\label{eq:cross_rho_ela1}
\end{equation}

where $V_0$ and $F(\tau)$ are the unit cell volume and the structure factor, respectively, with $\tau = 2\pi / d$ being the length of the reciprocal lattice vector relating to the lattice spacing $d$, $k=2\pi / \lambda$ is the modulus of the wavevector and $\lambda$ is the wavelength. Using the Miller indices ({\emph{hkl}}), Eq.~(\ref{eq:cross_rho_ela1}) can be rewritten as

\begin{equation}
\sigma_{coh, ela}(\lambda) = \frac{\lambda^2N}{2 V_0} \sum_{hkl} |F_{hkl}|^2 d_{hkl},
\end{equation}

Eq.~(\ref{eq:cross_rho_ela1}) is derived for a crystalline powder by averaging over a large number of crystallites with random isotropic orientation distribution. It hence integrates over all crystal orientations which fulfill, for a given wavelength $\lambda$, the Bragg condition 

\begin{equation}
n\lambda = 2d\sin\theta 
\label{eq:bragg_eq}
\end{equation}

In diffraction, the associated Debye-Scherrer cones are described by the delta function $\delta(1-\frac{\tau^2}{2k^2} - \cos\theta)$, which defines the characteristic cross section for powder-like samples with specific Bragg edge patterns in transmission.

For a single crystal, only one orientation has to be taken into account, and the elastic coherent scattering cross section changes to

\begin{equation}
\sigma_{coh, ela}(\tau, k) = \frac{(2\pi)^3 N}{2 V_0 k \tau}\sum_{\tau}|F(\tau)|^2 \delta({\tau^2 - 2k\tau\sin\theta})
\end{equation}

where the delta function represents the Bragg equation. The cross section is different from zero only when the Bragg condition is fulfilled, that is for the incoming wavelengths

\begin{equation}
\lambda = \frac{2\pi}{k} = \frac{4\pi}{\tau}\sin\theta
\label{eq:Bragg}
\end{equation}

Contrary to the powder case, where the Bragg edge pattern in the wavelength-dependent cross section is independent of the sample orientation, in the single crystal case the cross section displays discrete Bragg peaks at distinct wavelengths, which depend on the orientation of the crystal with respect to the beam. For textured materials with powder-type diffraction patterns, an angle-dependent additional factor, derived from the corresponding orientation distribution function ({\footnotesize{ODF}}) of all crystallites in the sample, has to be added \cite{santisteban2012texture}. 

In diffraction, a sample consisting of a large number of grains can be considered a powder when the beam size is much larger than the average grain size, while in transmission it is the spatial resolution that has to be much larger than the grain size. When the grain size is of the order of the spatial resolution, the sample is classified as a polycrystal, and the shape and orientation of the grains can in principle be reconstructed directly and in {\footnotesize{3D}} from a tomographic scan. In this case, the transmission signal can be expressed using a sum of discrete single crystal cross sections, and the number of observed Bragg peaks increases correspondingly to the number {\emph{n}} of grains intersected by the beam \cite{santisteban2001time}

\begin{equation}
\sigma_{coh, ela}(\lambda) = \sum_n \frac{(2\pi)^3 N}{2 V_0 k \tau}\sum_{hkl}|F_{hkl}|^2\delta(\tau^2 - 2k\tau \sin\theta_{n,hkl})
\label{eq:cross_rho_ela2}
\end{equation}

If the sample consists of both polycrystal regions and powder-like regions, the measured cross section will be a sum of the cross sections given by Eq.~(\ref{eq:cross_rho_ela1}) and Eq.~(\ref{eq:cross_rho_ela2}).

\begin{figure}
 \centering
 \includegraphics[width=1\textwidth]{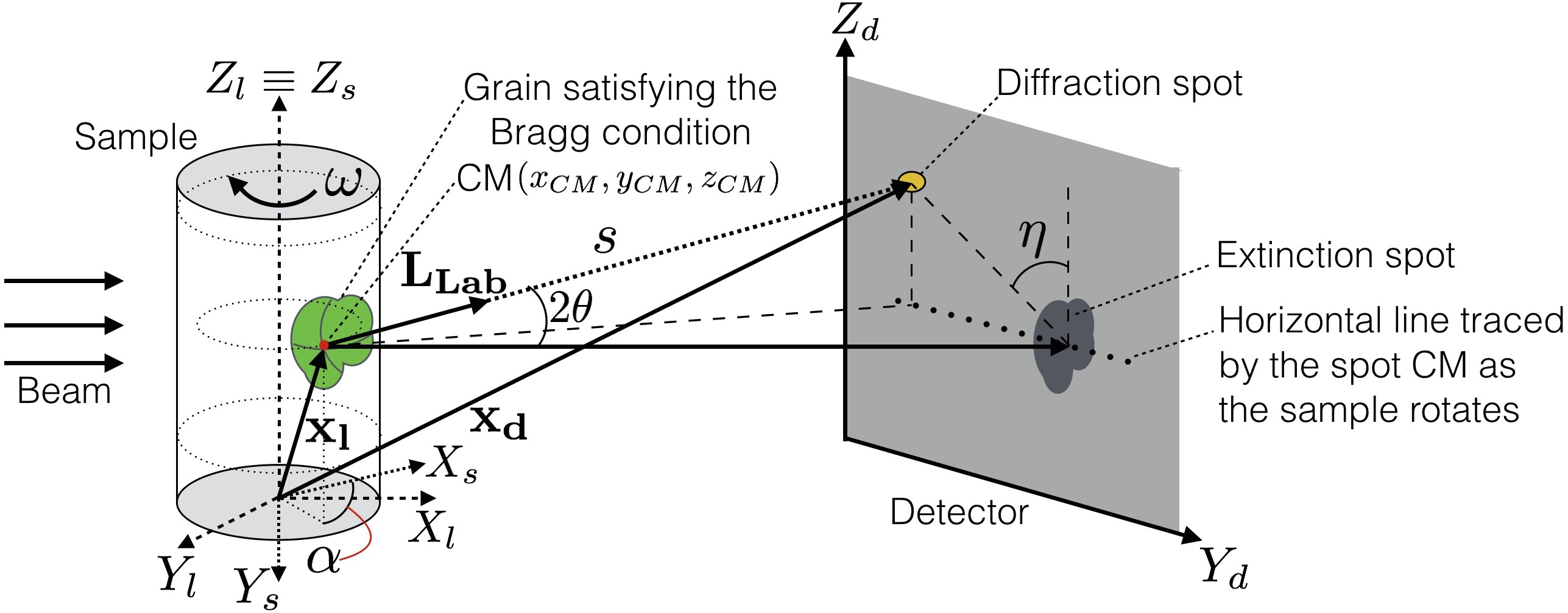}
 \caption{Experimental {\footnotesize{ToF 3DND}} setup. The sample is illuminated by a time-of-flight neutron beam and an imaging detector is installed in transmission geometry. When, at a sample rotation angle $\omega$, a grain satisfies the Bragg condition, a region of missing intensity ({\emph{extinction spot}}) is visible in transmission. The diffracted beam forms a bright {\emph{diffraction spot}}. In the plot, the laboratory, sample and detector reference systems are denoted by $(X_l, Y_l, Z_l)$, $(X_s, Y_s, Z_s)$ and $(Y_d, Z_d)$ respectively. While the {\footnotesize{3D}} shape of the grains can be reconstructed using extinction spots only, uniquely determining their orientation requires considering also the location of the diffraction spots. In real space, this position is determined by the direction of the diffraction vector ${\mathbf{L_{Lab}}}$, which connects the centre of mass ({\footnotesize{CM}}) of the grain, located by ${\mathbf{x_l}}$, with the centre of mass of the related diffraction spot. The relationship between ${\mathbf{x_l}}$ and the position, in the laboratory reference system, of the centre of mass ${\mathbf{x_d}}$ of the diffraction spot on the detector plane is ${\mathbf{x_d}} = {\mathbf{x_l}} + {\mathbf{L_{Lab}}} \cdot s$.}
 \label{fig:Detailed_geometry}
\end{figure}

When a grain in a polycrystalline sample satisfies the Bragg condition, the contribution to the attenuation signal is strong enough that missing intensity regions ({\emph{extinction spots}}) become visible in transmission mode with sufficient wavelength resolution (see Fig. \ref{fig:Detailed_geometry}). In the approach presented in this article, the {\footnotesize{3D}} shape and the orientation of the grains within a given sample are calculated from the distribution of the corresponding extinction and diffraction spots among the frames collected by the transmission detector at different rotation angles of the sample and wavelengths.

The extinction and diffraction spots are located by considering their signature in the transmission measurements, i.e. the ``Bragg peaks'' in the wavelength and angle dependent scattering cross section. Reconstruction of the individual grains in {\footnotesize{3D}} is possible by using a broad wavelength range, and rotating the specimen to perform a tomographic sampling. Furthermore, indexing and orientation mapping of the individual grains can be determined by correlation of the extinction spots from the near field detector data. Good wavelength resolution over the broad wavelength band provides a large number of peaks, each contributing to the spatially resolved cross section. 

\section*{Experimental}

\subsection*{Setup at J-PARC}

To collect information on the shape and location of the extinction spots, and on the wavelength where they are collected, a time-of-flight neutron beam is used \cite{anderson2009neutron}. With such a beam, the kinetic energy (and thus the wavelength) of a traveling neutron can be calculated by the time it takes him to fly between two fixed points whose distance is known.

The neutron experiments were conducted at beamline {\footnotesize{BL18}}, the single crystal {\footnotesize{ToF}} diffractometer {\footnotesize{SENJU}},  of the {\footnotesize{MLF}} of {\footnotesize{J-PARC}}, a short pulse neutron source \cite{nagamiya2012introduction, Oikawa_2014}. Data were acquired with the source operating respectively at 300 kW (Fe sample) and at 200 kW (Co-Ni-Ga sample). The instrument was chosen for its good wavelength resolution and large bandwidth. The short pulse combined with a moderator-to-detector distance of 34.8 m and the 25 Hz source frequency enables a wavelength resolution of 0.3\% over a range of 4 {\AA} (between 0.4 and 4.4 {\AA}), selected by two bandwidth choppers. 

{\footnotesize{ToF 3DND}} measurements require installing a high resolution {\footnotesize{ToF}} imaging detector \cite{Anton_2013}, referred to as the {\emph{transmission detector}}, placed a few centimeters behind the sample. The detector employs a microchannel plate ({\footnotesize{MCP}}) with an active area of 28$\times$28 mm$^2$, corresponding to 512$\times$512 pixels, each with 55 $\mu$m size. The detector, suited for count rates of $> 10$ MHz in ToF mode, and up to GHz in counting mode, provides approximately 100 $\mu$m intrinsic spatial resolution, paired with $<1 \mu$s time resolution, which makes it well suited for a wide range of neutron {\footnotesize{ToF}} imaging applications \cite{strobl2015quantitative, tremsin2012high, tremsin2015spatially}. The transmission data were recorded using the Pixelman software package \cite{turecek2011pixelman}. The experimental geometry is sketched in Fig. \ref{fig:Detailed_geometry}.

As a reference sample we used an ultrapure (99.98\% purity) iron cylinder, 5 cm long and 1 cm in diameter. The iron sample was obtained from a rod purchased from Goodfellow Cambridge
Ltd. (Huntingdon, England), cold worked, cut, vacuum-sealed in quartz-glass capsules and heat treated at 900$^{\circ}$ C for 20 days to grow grains with size in the hundred of microns to millimeter range. The {\footnotesize{ToF 3DND}} algorithms were cross-checked using a cobalt-nickel-gallium (Co-Ni-Ga) sample, consisting of two stacked single-crystal cubes with a size of 4 mm \cite{vollmer2015damage}. With its very simple arrangement of the grains, the Co-Ni-Ga sample serves as a good validation of the indexing procedure.

The transmission detector recorded projections over 54 rotation angles over 180$^\circ$, with an exposure time of 1 hour per projection. For a given projection, data were recorded in {\footnotesize{ToF}} histogram mode\cite{ohhara_2016} relative to the neutron trigger pulse with the full {\footnotesize{ToF}} range of 36.35 ms, split into 2423 bins of 12.8 $\mu$s. Measures were complemented by acquiring datasets with no sample in the beam ({\emph{open beam}}), both before and after imaging the sample. Due to the long sample exposure, variations of the detection efficiency are expected for the transmission detector.

\subsection*{Electron backscatter diffraction}

After being studied using {\footnotesize{ToF 3DND}}, the Fe sample was sliced longitudinally and radially into four pieces and the longitudinal surfaces were polished and imaged using electron backscatter diffraction ({\footnotesize{EBSD}}) in a scanning electron microscope ({\footnotesize{SEM}}) \cite{EBSD_book}. The measurements were performed on a {\footnotesize{FEI}} (part of Thermo Fisher Scientific, Waltham, {\footnotesize{MA}}, {\footnotesize{USA}}) Nova NanoLab 600 microscope equipped with a Bruker (Billerica, {\footnotesize{MA}}, {\footnotesize{USA}}) $e^-$Flash\textsuperscript{HD} {\footnotesize{EBSD}} detector. The data analysis was performed using CrystAlign by Bruker (Billerica, {\footnotesize{MA}}, {\footnotesize{USA}}), and {\footnotesize{TSL OIM}} Analysis\textsuperscript{TM} software by {\footnotesize{EDAX}} (Mahwah, {\footnotesize{NJ}}, {\footnotesize{USA}}).

\section*{Methods}

\subsection*{Grain shape and juxtaposition}

The approach used to reconstruct the {\footnotesize{3D}} shape and juxtaposition of the grains within a sample combines ordinary and new procedures. While the pre-processing and {\footnotesize{3D}} reconstruction procedures are fairly standard \cite{strobl2009advances}, most of the data processing solutions are tailored for {\footnotesize{ToF 3DND}}. The presented approach, developed for the Fe sample, was used in a simplified version for the Co-Ni-Ga sample. 

Data analysis algorithms were developed using {\footnotesize{MATLAB}} and Image Processing Toolbox Release 2016a, The MathWorks, Inc. (Natick, {\footnotesize{MA}}, {\footnotesize{USA}}). Basic image processing was performed using {\footnotesize{FIJI}} \cite{schindelin2012fiji} and Adobe Photoshop by Adobe System (San Jose, {\footnotesize{CA}}, {\footnotesize{USA}}). Grain structures were rendered in {\footnotesize{3D}} using Paraview \cite{ahrens200536}.

The image datasets acquired by the transmission detector were pre-processed using a dead-time correction algorithm \cite{Anton_2014}, and normalized by the relative open beam and using a rolling median. Pre-processed images were then refined using the multiplicative rotational filter Murofi (to be published in a separate article), specifically developed to segment weak extinction spots with varying shape and intensity in transmission images. For a given grain, the heterogeneity of the shapes is due to the fact that, for a given projection, the corresponding extinction spots are observed in a number of wavelength intervals, with continuously changing shape around a Bragg peak. This spreading, visible with the chosen temporal resolution, is ultimately due to resolution effects and to the mosaicity of the grains, estimated from {\footnotesize{EBSD}} to be less than 2$^{\circ}$. In the present work, we therefore integrate over the full Bragg peak. Resolving the local mosaicity is a natural extension of the methodology.

\begin{figure}
\includegraphics[width=0.8\textwidth]{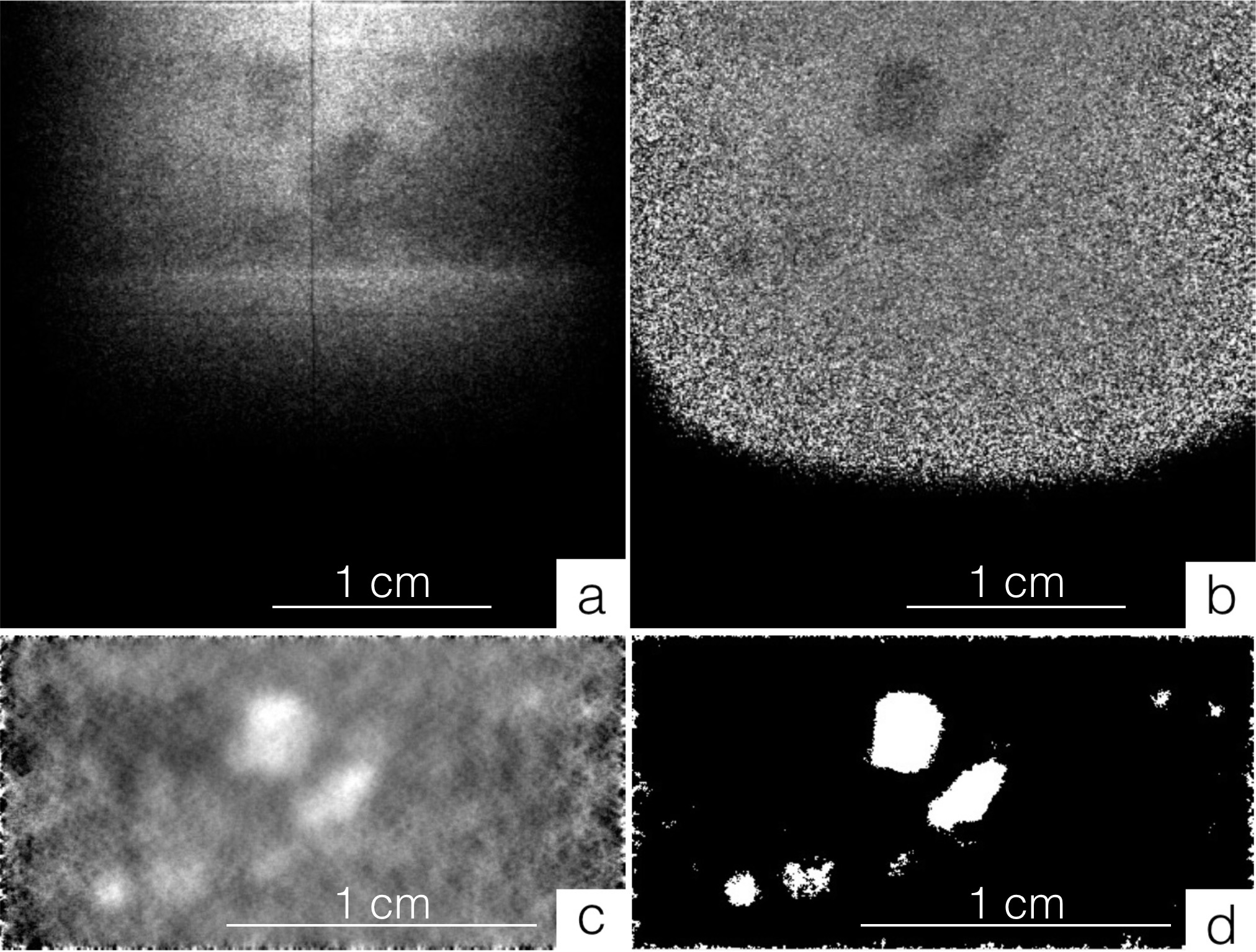}
\centering
\caption{Image processing for extinction spots segmentation. The multiplicative rotational filter Murofi (to be published in a separate article) was developed to locate extinction spots in transmission images collected using a {\footnotesize{ToF}} beam. {\textbf{(a)}} Raw image collected studying the Fe sample. {\textbf{(b)}} {\textbf{a}} after pre-processing. {\textbf{(c)}} {\textbf{b}} after being filtered using Murofi. {\textbf{(d)}} {\textbf{c}} binarized using a threshold value. In {\textbf{c}} and {\textbf{d}}, only the sample region is shown.}
\label{fig:Murofi_gs_bw}
\end{figure}

After being filtered and binarized (see Fig. \ref{fig:Murofi_gs_bw}), the projections are processed to select the extinction spots. As a result of the spreading over {\footnotesize{ToF}}, spots related to different grains may overlap, which makes small grains harder to identify and to reconstruct than big ones, because their extinction spots are more easily covered. For a sample composed of both small and big grains, such as the Fe sample, extinction spots are processed in separate groups, each covering a different range of area ({\emph{A}}) values. In the Fe case, extinction spots are divided in four groups, depending on their area: between 100 and 500 pixels, between 100 and 1000 pixels, between 100 and 2000 pixels, and greater than 1000 pixels. To avoid missing grains, there is a partial overlap between the different area intervals. Grouping the extinction spots is not necessary when the investigated sample is made of grains with similar size, as in the Co-Ni-Ga case.

\subsubsection*{Grain sorting}
\label{subs:grain_sorting}

For each projection, it is necessary to group extinction spots recorded at different wavelengths that are related to the same grain. Spots are grouped by similarity, using a criterion built on morphological operations (see Supplementary Fig. S1) \cite{russ2016image, sonka2014image}. Similar spots are combined and possible duplicates (with centre of mass very close) merged. For each grain, the final combination of the related extinction spots is considered as the best estimate of its projected shape at a given rotation angle.

When a sample is illuminated by a neutron beam and rotated around its vertical axis, the centre of mass of the extinction spots related to a given grain (corresponding, roughly, to the projection of the centre of mass of the grain) moves along a horizontal line on the detector plane $y_d z_d$ (see Fig. \ref{fig:Detailed_geometry}). This trajectory corresponds to a sinusoid in the $y_d z_d \omega$-space, with $\omega$ being the rotation angle. Extinction spots are divided in groups considering the distribution of their centers of mass in the $y_d z_d \omega$-space: in that space, spots related to the same grain have centre of mass distributed around the same sinusoid. On the detector plane, the sinusoids are straight lines parallel to the $Y_d$-axis. To account for different sinusoids, a rolling window sweeps the $Z_d$-axis. For each interval, the Hough transform\cite{haidekker2011advanced, hough1959machine} is used to calculate the corresponding sinusoids in the $y_d \omega$-plane.

For a given grain with centre of mass $\mathbf{r_0}=\left(\begin{smallmatrix}x_{CM} \\ y_{CM} \\ z_{CM} \end{smallmatrix}\right)$ in the sample reference system, the centre of mass $\mathbf{r_d} =\left(\begin{smallmatrix}x_d \\ y_d \\ z_d \end{smallmatrix}\right)$ in the detector reference system is given by $\mathbf{r_d} = \Omega\mathbf{r_0}$, where $\Omega$ is the left-handed rotation matrix around the $z$-axis by an angle $\omega$ 

\begin{equation}
  \Omega =
  \left(
  \begin{array}{ccc}
    \cos\omega & \sin\omega & 0 \\
    -\sin\omega & \cos\omega & 0 \\
    0 & 0 & 1
  \end{array}
  \right)
\end{equation}

which is the transformation from the sample reference system to the laboratory reference system. 

As the sample rotates, the curve described on the detector surface by the centre of mass of the grain is defined by the equations

\begin{align}
  y_d & = - x_{CM}\sin{\omega} + y_{CM}\cos{\omega} = R \cdot \cos (\omega + \alpha)\\
  z_d & = z_{CM}
  \label{eq:yz_CM}
\end{align}

with $R = \sqrt{x_{CM}^2 + y_{CM}^2}$. $\omega$ and $\alpha$ are, respectively, the sample rotation angle at which a given projection has been collected and the angle, in the sample reference system, describing the position of the centre of mass of the grain in the $x_s y_s$-plane using polar coordinates. To group the centre of mass points, we selected a region around the calculated $y_d = R \cdot \cos (\omega + \alpha)$ curves. The parameters $R$ and $\alpha$ are returned by the Hough transform,  which represents points in the $\omega y_d$-space as curves in the  $R \alpha$-space. 

\subsubsection*{3D reconstruction}

Once the extinction spots related to a given grain are grouped, the {\footnotesize{3D}} shape of the grain can be reconstructed by back-projection. Different grain shapes are reconstructed separately and then assembled in a unique volume. If the extinction spots are divided in a number of size intervals and separately processed, several partial reconstructions are obtained. These are combined in a final reconstruction using a Russian-doll-like approach: if a grain is contained within a larger grain, it is considered part of it. If two grains from different reconstructions partially overlap, they are considered as a single entity if the centre of mass of one is positioned inside the volume of the other. 

The top of the Fe sample contains a large powder-like region made of crystallites with random orientations, probably due to oxidation effects of the oxygen trapped when sealing the quartz capsules. While the grains give transmission contrast only for certain $(\omega, \lambda)$ combinations, the powder-like region has a fingerprint that is recorded at all $\omega$  and $\lambda$: being formed by randomly oriented crystallites, for each $(\omega, \lambda)$ combination there is a number of crystallites which satisfy the Bragg condition. As a consequence, the powder-like region gives a contrast that cannot be given by the grains, and its reconstruction complements the reconstruction of the grains.  

\begin{figure}
\centering
\includegraphics[width=0.9\textwidth]{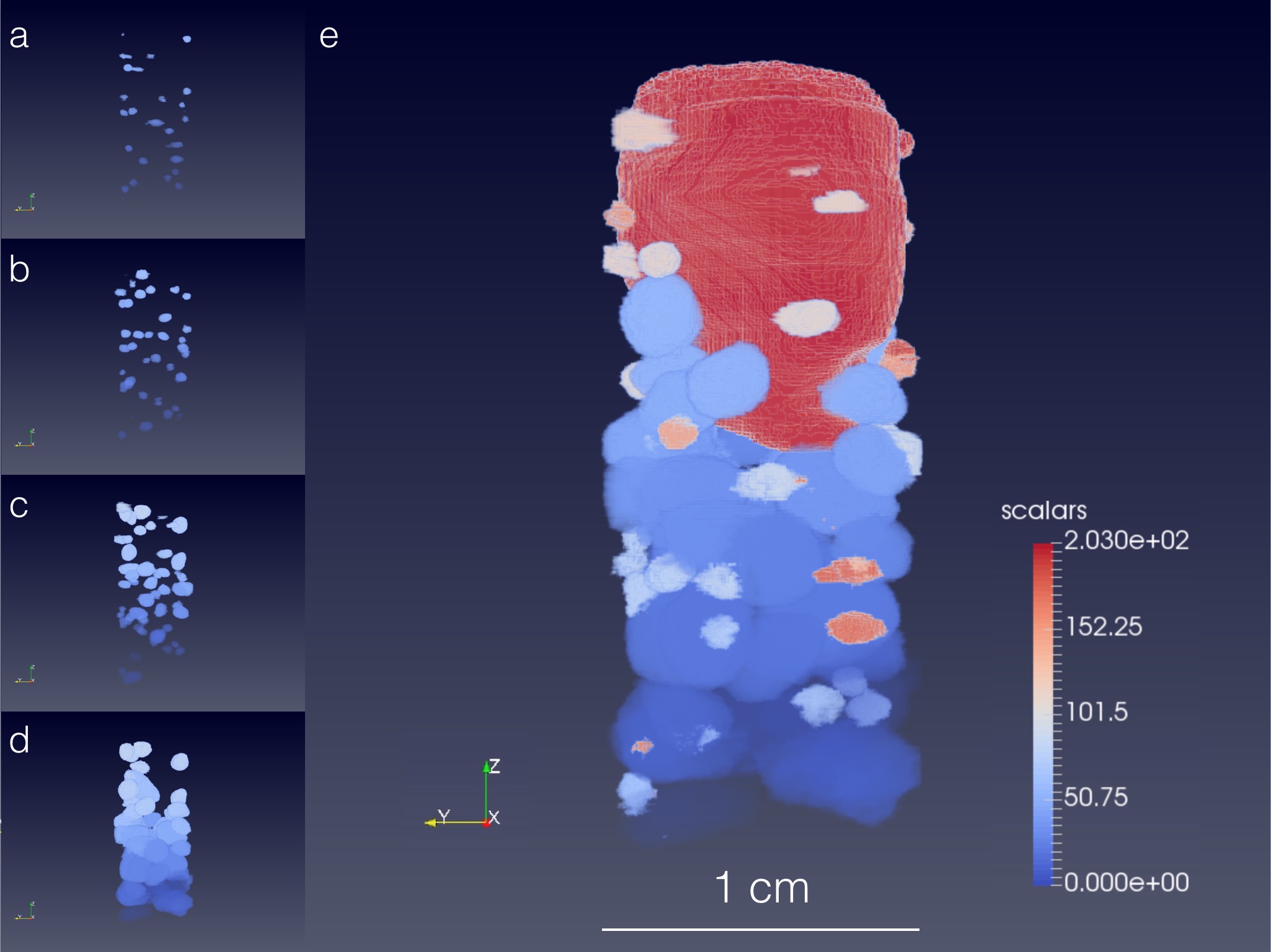}
\caption{Partial and final reconstructions of the Fe sample investigated at {\footnotesize{SENJU}}. {\textbf{(a-d)}} To reconstruct both small and big grains, extinction spots were divided in four size intervals and processed separately. Four {\footnotesize{3D}} reconstructions are obtained by backprojection: {\textbf{a}} for spots with an area between 100 and 500 pixels, {\textbf{b}} for spots with an area between 100 and 1000 pixels, {\textbf{c}} for spots with an area between 100 and 2000 pixels, and {\textbf{d}} for spots with an area greater than 1000 pixels. In each reconstruction, different grains are shown in different colors. {\textbf{(e)}} The final reconstruction, obtained by combining the partial reconstructions, consists of 108 grains, each represented in a different color. The upper part of the sample includes a large powder-like region, probably due to oxidation effects occurred during the sample heating treatment. The powder-like region is reconstructed using an inverse Radon transform.}
\label{fig:Final_reconstr}
\end{figure}

\begin{figure}
\centering
\includegraphics[width=1\textwidth]{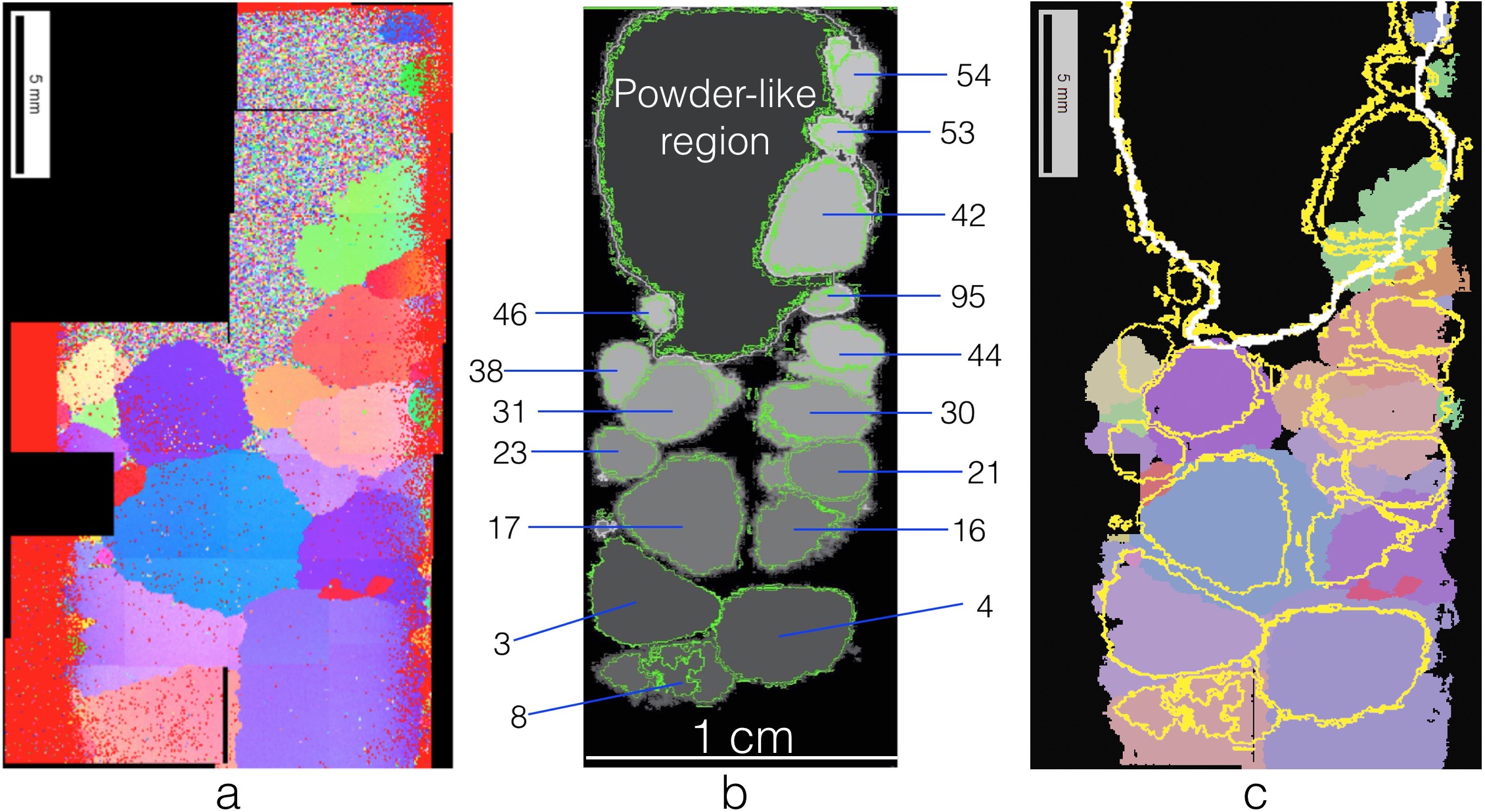}
\caption{Validation of the {\footnotesize{ToF 3DND}} shape reconstruction algorithms. A vertical slice of the {\footnotesize{3D}} reconstruction of the Fe sample is compared with the {\footnotesize{2D}} map obtained by electron backscattering diffraction ({\footnotesize{EBSD}}) after cutting the sample in two both longitudinally and radially. {\textbf{(a)}} {\footnotesize{EBSD}} map of a vertical sample slice (half length), with different orientations (and thus grains) shown in different colors. A powder-like region is visible at the top-right of the slice. {\textbf{(b)}} Corresponding slice from the {\footnotesize{ToF 3DND}} reconstruction, where the powder-like volume was added in a way to preserve the shape of the grains. The {\footnotesize{ID}} number of each grain is indicated. {\textbf{(c)}} {\footnotesize{EBSD}} map with, superimposed, the perimeter of the grains from the {\footnotesize{3DND}} reconstruction. The outline of the grains from {\textbf{b}} is plotted in yellow and, in the upper part, the outline of the powder-like region, reconstructed without preserving the shape of the grains, is shown in white. While the {\footnotesize{EBSD}} map in {\textbf{a}} shows the raw data, the one in {\textbf{c}} shows, for each grain, the average orientation. In {\textbf{c}}, both powder-like and non-measured regions are shown in black.}
\label{fig:EBSD_3DND}
\end{figure}

The powder-like volume is reconstructed by adding, for each projection, all normalized images, rescaling the sum and applying an inverse Radon transform \cite{Radon_paper, radon1986determination}. The resulting images are then combined in a {\footnotesize{3D}} volume by backprojection. The powder-like region is included in the final reconstruction in a way which preserves the shape of the grains. The uncertainty in the reconstruction of the powder-like volume is due to the limited number of considered projections (48). The partial and final reconstructions of the iron sample are shown in Fig. \ref{fig:Final_reconstr}. The reconstructed powder-like region well fits with the portion of the reconstructed sample containing no grains and it confirms the powder-like region observed in the {\footnotesize{EBSD}} measurements (see Fig. \ref{fig:EBSD_3DND}).

\subsection*{Grain orientation}
\label{sec:indexing}

The procedure developed to calculate the  crystallographic orientation of the grains ({\emph{indexing}}) is based on fitting the distribution of the center of mass of the extinction spots in the $\omega\lambda$-plane.

\begin{figure}
\centering
\includegraphics[width=1\textwidth]{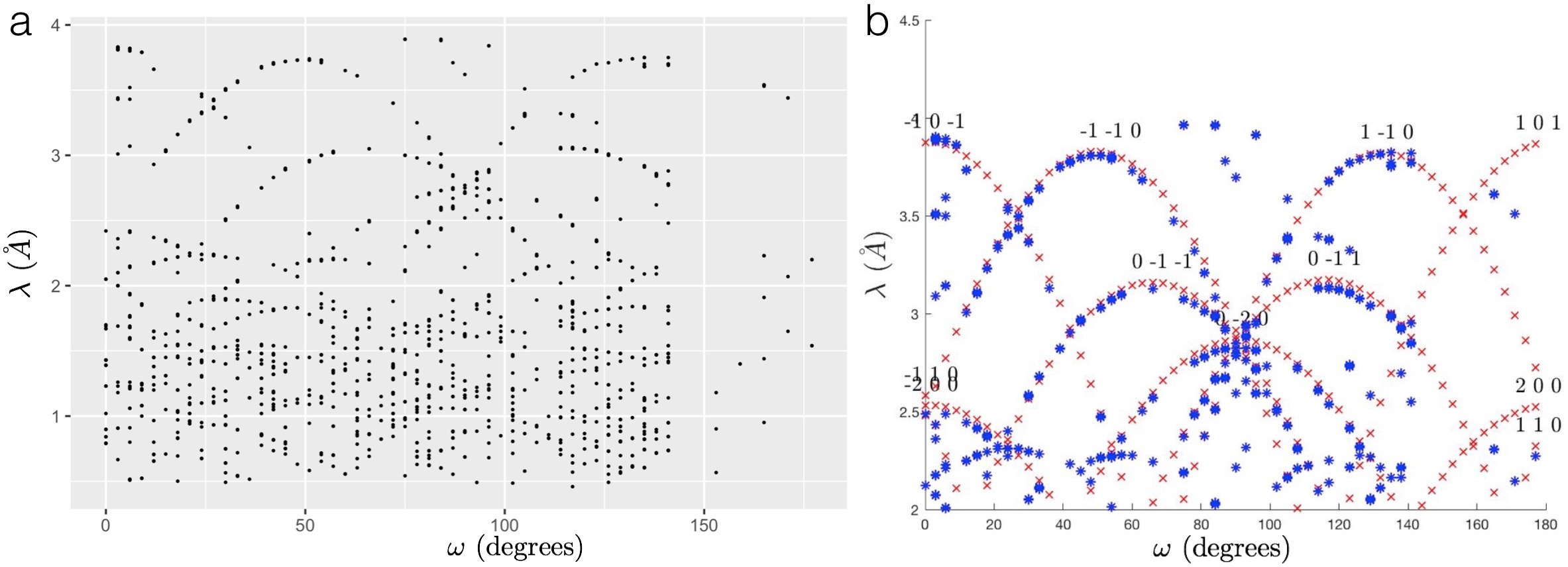}
\caption{Grain indexing based on extinction spots distribution. The images show the raw ({\textbf{a}}) and fitted ({\textbf{b}}) distribution, in the $\omega\lambda$-plane, of the extinction spots belonging to one of the grains in the investigated iron sample. $\omega$ is the sample rotation angle and $\lambda$ is the centre of the neutron wavelength interval in which a given extinction spot is visible. ({\textbf{b}}) Determining the equation of the curves along which the points are distributed is the first step to find the orientation of the corresponding grain. Different curves are related to different {\emph{hkl}} families.}
\label{fig:OL_raw_and_fit}
\end{figure}

In the $\omega\lambda$-plane, extinction spots are distributed along multiple curves (see Fig. \ref{fig:OL_raw_and_fit}). Such a distribution can be fitted by using a forward model which calculates the curves in the $\omega\lambda$-space. The grain orientation is the orientation that minimizes the distance of the curves from the experimental values. The starting equations are Eq. (\ref{eq:Bragg}), Bragg's law, and the diffraction equation \cite{poulsen2001three}

\begin{equation}
  \textbf{G} = \frac{d}{2\pi}\Omega \text{\emph{U}}\mathcal{B}\textbf{h} 
  \label{eq:fund_diff_eq}
\end{equation}

where $|\textbf{G}| = 1$, $d = \lambda/2\pi$ is the spacing between the lattice planes, $\Omega$ is the left-hand rotation matrix, $\text{\emph{U}}$ is the orientation matrix, $\textbf{h} = \left(\begin{smallmatrix}h \\ k \\ l\end{smallmatrix}\right)$ and $\mathcal{B}$ is the matrix transforming the {\emph{hkl}} lattice into reciprocal space. 

Combining Eq.~(\ref{eq:bragg_eq}) and Eq.~(\ref{eq:fund_diff_eq}), it is possible to write the functions of the fitting curves

\begin{equation}
  \lambda(\omega) = -\frac{4\pi}{|{\mathcal{B}\textbf{h}}|^2}(A \cos\omega + B \sin\omega)
  \label{eq:final_OL}
\end{equation}

%where the coefficients $A$ and $B$, constant for a given {\emph{hkl}} family, are functions of the elements of $\text{\emph{U}}$ and of $\mathcal{B}\textbf{h}$. For more details, see Supplementary Sec. \ref{sec:suppl_calc_LO}. 

where the coefficients $A$ and $B$, constant for a given {\emph{hkl}} family, are functions of the elements of $\text{\emph{U}}$ and of $\mathcal{B}\textbf{h}$. For more details, see Supplementary Sec. S2. 

\subsubsection*{Indexing procedure}

Eq.~(\ref{eq:final_OL}) returns, for a number of possible orientations $U_i$, the relative functions $\lambda_i(\omega)$ for a number of chosen $hkl$ sets. Different $\lambda_i(\omega)$ curves are tested to fit the distribution of points in the $\omega\lambda$-space, applying a forward model, and the orientation which best fits the experimental values is chosen. The procedure is outlined in detail in the Supplementary Sec. S3. An efficient way to test all possible orientations is to sample the fundamental zone of the Rodrigues space, which for the cubic crystal system is a truncated cube of size $2\cdot(\sqrt{2}-1)$ \cite{frank1988orientation, Morawiec_1996, he2007representation}. The selected orientation can be iteratively refined by sampling, with smaller steps, a selected volume in the fundamental zone built around the previously identified orientation vector.

%Eq.~(\ref{eq:final_OL}) returns, for a number of possible orientations $U_i$, the relative functions $\lambda_i(\omega)$ for a number of chosen $hkl$ sets. Different $\lambda_i(\omega)$ curves are tested to fit the distribution of points in the $\omega\lambda$-space, applying a forward model, and the orientation which best fits the experimental values is chosen. The procedure is outlined in detail in the Supplementary Sec. \ref{sec:suppl_index_proc}. An efficient way to test all possible orientations is to sample the fundamental zone of the Rodrigues space, which for the cubic crystal system is a truncated cube of size $2\cdot(\sqrt{2}-1)$ \cite{frank1988orientation, Morawiec_1996, he2007representation}. The selected orientation can be iteratively refined by sampling, with smaller steps, a selected volume in the fundamental zone built around the previously identified orientation vector. 

Since data are acquired using only one rotation axis, a two-fold ambiguity exists in the determination of the crystallographic orientation (see Supplementary Sec. S4). This ambiguity can be solved by considering the position of the diffraction spots that are recorded on the transmission detector in forward direction.

%Since data are acquired using only one rotation axis, a two-fold ambiguity exists in the determination of the crystallographic orientation (see Supplementary Sec. \ref{sec:suppl_U_CUC}). This ambiguity can be solved by considering the position of the diffraction spots that are recorded on the transmission detector in forward direction.

Let us introduce the two vectors $\mathbf{G^r}$ and $\mathbf{L}$. $\mathbf{G^r}$ is the reciprocal vector and $\mathbf{L}$ is the direction, in the laboratory reference system, of the diffraction vector in real space (see Fig. \ref{fig:Detailed_geometry}). The vectors are defined as \cite{schmidt2014grainspotter}

\begin{align}
  \mathbf{G^r} 		& = \Omega^{-1} \mathbf{G} \\
  \mathbf{L_{Lab}} 	& = \Omega \mathbf{L} = \Omega (2\mathbf{G^r} + \Omega^{-1}\vert_1 )
  \label{eq:L_Gr}
\end{align}

with $\textbf{G} = \frac{\lambda}{4\pi}\Omega \text{\emph{U}}\mathcal{B}\textbf{h}$, as presented in Eq.~(\ref{eq:fund_diff_eq}). 

\begin{figure}
\centering
\includegraphics[width=1.0\textwidth]{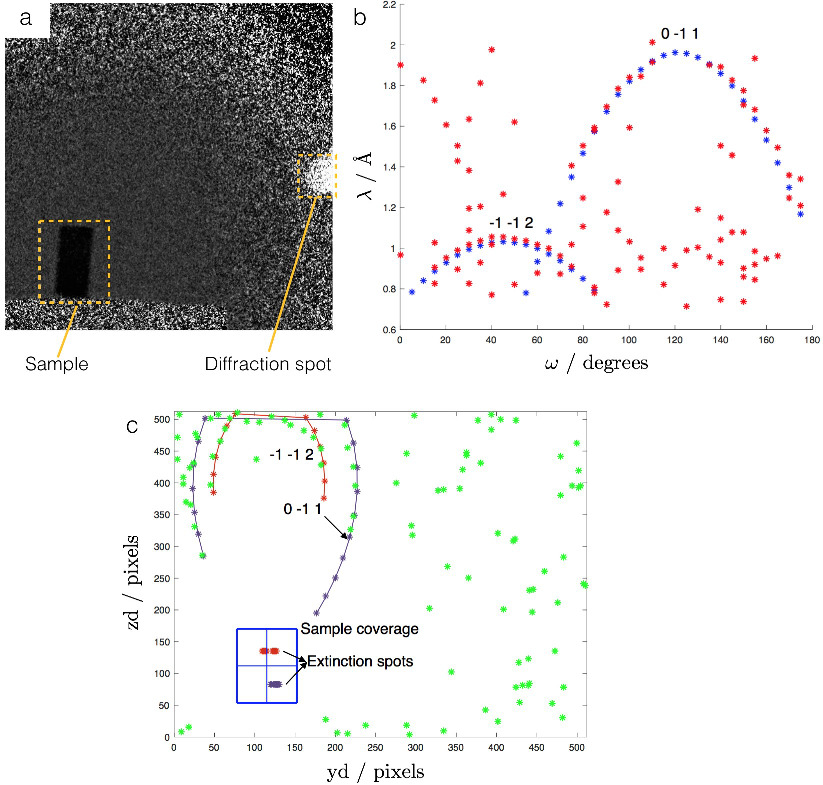}
\caption{Determination of the grain orientations using diffraction spots. For the Co-Ni-Ga sample, it was possible to uniquely define the orientation of the grains by using the position of the related diffraction spots on the detector. {\textbf{(a)}} Image recorded illuminating the Co-Ni-Ga sample, with a visible diffraction spot. {\textbf{(b)}} Distribution of the detected extinction spots (in red) as a function of the sample rotation angle $\omega$ and of the wavelength $\lambda$. The distribution was well fitted for the {\emph{hkl}} values -1-12 and 0-11, each related to a different grain. The points of the fitting curves are shown in blue. {\textbf{(c)}} Using the predicted positions of the diffraction spots on the {\footnotesize{MCP}} detector (determined by $\mathbf{x_d}$, see Fig. \ref{fig:Detailed_geometry}), it was possible to fit the experimental values (in green) and thus to uniquely define the orientation of the grains. The fitting curves are relative to the {\emph{hkl}} values 1-12 and 0-11.}
\label{fig:fitting_curves_bamboo}
\end{figure}

For a given grain with known centre of mass position, $\mathbf{L_{Lab}}$ dictates the position of the related diffraction spot on the detector plane as a function of $\text{\emph{U}}$ and {\emph{hkl}} (see Fig. \ref{fig:Detailed_geometry}). Considering the limited portion of diffraction spots collected by the transmission detector, the orientation of a grain can be uniquely determined by fitting the experimental data using $\mathbf{L_{Lab}}$ (see Fig. \ref{fig:fitting_curves_bamboo}).

\section*{Results}

Fig. \ref{fig:Final_reconstr} shows, in {\footnotesize{3D}}, the internal structure of the Fe sample investigated using {\footnotesize{ToF 3DND}}. The sample consists of 108 grains, with volume ranging from $1.49\cdot 10^{-2}$ $\text{mm}^3$ to 59.1 $\text{mm}^3$, and a large powder-like volume. The data acquisition setup did not allow to locate enough diffraction spots to uniquely determine the orientation of the grains. This limitation can be overcome by partially masking the incoming beam, to illuminate the sample only and reduce the background from the direct beam, or by introducing a second rotation axis perpendicular to the $Z_s$-axis. 

When studying the Co-Ni-Ga sample it was possible to localize the collected diffraction spots and thus to uniquely determine the orientation of the two grains the sample is made of. To check the validity of the grain orientations, they were compared with the results of {\footnotesize{EBSD}}. The agreement between the different methods of calculating the grain orientation is measured by considering the misorientation between the grains ($\phi_{min}$), defined as

\begin{equation}
\phi_{min} = \min_{i}\bigg[\arccos\bigg(\frac{Tr(U_1 E_i U_2')-1}{2}\bigg)\bigg]
\end{equation}

where $U_1$ and $U_2$ are orientation matrices related to different grains, \emph{Tr} is the matrix trace and $E_i$ are the space group generators, i.e. the symmetry operators of a given crystalline structure \cite{rupp2009biomolecular}. From the {\footnotesize{ToF 3DND}} data, the misorientation is 40$^{\circ}$, and from the {\footnotesize{EBSD}} data it is 40.5$^{\circ}$, using Bunge’s convention \cite{maitland2007electron}. Considering that the uncertainty from the {\footnotesize{EBSD}} measurements is between 1$^{\circ}$ and 2$^{\circ}$, the agreement between the two misorientation values is good. This validates the developed indexing method.

%The algorithms developed to reconstruct the shape of the grains were cross-checked by comparing the outline of the grains returned by {\emph{post-mortem}} {\footnotesize{EBSD}} with the perimeter of the grains in the related vertical slice of the {\footnotesize{3D}} sample reconstruction, as shown in Fig. \ref{fig:EBSD_3DND}. The two maps show good agreement. In the {\footnotesize{EBSD}} grain map, part of the sample shows a powder-like structure, which corresponds to the powder-like region added by conventional reconstruction to the {\footnotesize{ToF 3DND}} grain map. Moreover, the reconstruction algorithms developed for the Fe sample are also verified by reconstructing the shape and juxtaposition of the two grains composing the Co-Ni-Ga sample. Supplementary Fig. \ref{sfig:Bamboo_sample} shows two grain maps of the Co-Ni-Ga sample, respectively obtained using {\footnotesize{ToF 3DND}} and {\footnotesize{EBSD}}.

The algorithms developed to reconstruct the shape of the grains were cross-checked by comparing the outline of the grains returned by {\emph{post-mortem}} {\footnotesize{EBSD}} with the perimeter of the grains in the related vertical slice of the {\footnotesize{3D}} sample reconstruction, as shown in Fig. \ref{fig:EBSD_3DND}. The two maps show good agreement. In the {\footnotesize{EBSD}} grain map, part of the sample shows a powder-like structure, which corresponds to the powder-like region added by conventional reconstruction to the {\footnotesize{ToF 3DND}} grain map. Moreover, the reconstruction algorithms developed for the Fe sample are also verified by reconstructing the shape and juxtaposition of the two grains composing the Co-Ni-Ga sample. Supplementary Fig. S3 shows two grain maps of the Co-Ni-Ga sample, respectively obtained using {\footnotesize{ToF 3DND}} and {\footnotesize{EBSD}}.

\section*{Discussion and conclusion}

{\footnotesize{ToF 3DND}} is a new, nondestructive, {\footnotesize{3D}} neutron imaging technique that can return information on both the shape and orientation of the grains in polycrystalline materials. {\footnotesize{ToF 3DND}} is particularly adapted to studies of samples that cannot be studied by {\footnotesize{X}}-ray methods due, for example, to large dimensions and high {\footnotesize{X}}-ray attenuation, e.g., centimeter-sized metallic specimens containing grains in the hundreds of microns to millimeter scale. {\footnotesize{ToF 3DND}} also has advantages over another recent neutron-diffraction based imaging approach, namely n{\footnotesize{DCT}}, in that it has less limitations in terms of the number and size of the grains (n{\footnotesize{DCT}} is limited by diffraction spots overlapping and blurring, which set a minimum grain size of 1 mm for a mosaicity of 0.1-0.2$^{\circ}$)\cite{peetermans2014cold}. In fact, {\footnotesize{ToF 3DND}} allows 10 times more grains to be resolved than any earlier {\footnotesize{3D}} neutron diffraction approach and with better spatial resolution. This performance is possible by adding relatively little additional equipment to a time-of-flight neutron beamline, involving only an imaging detector with high spatial and temporal resolution.

In this article we have outlined the {\footnotesize{ToF 3DND}} technique and demonstrated its application with the successful reconstruction of 108 grains in a sample of iron, which was validated by comparison with a grain map obtained by {\footnotesize{EBSD}}. Furthermore, we uniquely determined the orientation of two grains composing a Co-Ni-Ga sample, as confirmed by {\footnotesize{EBSD}}. These developments open up possibilities to investigate other polycrystalline materials that could not be studied with previously available techniques. By correlating information from {\footnotesize{ToF 3DND}} and from recently developed tools for microstructure analysis of mosaic crystals \cite{malamud2016full}, it will be possible to visualize in {\footnotesize{3D}} how lattice parameters, mosaicity, extinction factors and crystal orientation vary across a sample. {\footnotesize{ToF 3DND}} could also be used to study geological samples or to investigate the formation of meteorites. As a non-destructive approach, {\footnotesize{ToF 3DND}} will enable in-situ, time-resolved studies of multi-grain structures under external forces (e.g. loading) and might be used to analyze force chains in granular materials \cite{peters2005characterization,wensrich2014force, hall2015three, hurley2016quantifying}. Combining {\footnotesize{ToF 3DND}} with neutron Bragg-edge tomography or polarized neutron imaging of magnetic domains \cite{woracek2015neutron, kardjilov2008three} would pave the way for a new generation of multidimensional maps showing, on top of the grain shape and orientation, how phase and magnetic domains are arranged in the sample. 

An extension of the described {\footnotesize{ToF 3DND}} approach is to combine the information from a transmission detector with data from diffraction detectors. If both data sets are available, the grain orientations can be determined from the diffraction data and used to select the transmission data frames containing the associated extinction spots. In this way, it should be possible to reconstruct small grains that are difficult to observe using the transmission detector alone. Therefore, by including both transmission and diffraction data, {\footnotesize{ToF 3DND}} has the potential to reconstruct grain maps for samples made of thousands of grains. This approach will be presented in a forthcoming paper. 

%\bibliography{3DND}

\section*{Acknowledgements}

%3DND has been developed basing on data collected using the {\footnotesize{SENJU}} diffractometer, installed at beamline {\footnotesize{BL18}} of the Public Neutron beam Facility of {\footnotesize{J-PARC}} (Tokai, Japan). % Required by J-PARC
A.C., E.B.K, M.Sa., P.M.L. and S.S.  thank {\footnotesize{DANSCATT}} for financial support. The PhD project of A.C. was partially financed by {\footnotesize{ESS}}. The neutron experiments at the Materials and Life Science Experimental Facility of the {\footnotesize{J-PARC}} were performed under a user program ({\footnotesize{2014A0070}} and {\footnotesize{2014B0132}}). P.K. and T.N. thank the Deutsche Forschungsgemeinschaft ({\footnotesize{DFG}}), contract {\footnotesize{NI1327/3-2}}.

\section*{Author contributions statement}

The experiment was conceived by S.S. and M.St.  The experiment was prepared by A.C., S.S., E.B.K., P.W., R.K., A.S. and S.H. Measurements at {\footnotesize{J-PARC}} were conducted by A.C., S.S., A.T., R.K., M.St., E.B.K., P.M.L., T.S., T.H., T.M., M.Sa., P.K., P.M.K. and P.M.L. The samples were prepared and characterized {\emph{post-mortem}} by A.C., A.B.d.S.F., S.I., P.K., P.M.K. and T.N. Algorithm development as well as data analysis was done by A.C. and S.S. The manuscript was prepared by A.C., M.St. and S.S. All authors reviewed the manuscript.  

%\clearpage

\section*{Supporting Information}.

%These commands reset the figure counter and add "S" to the figure caption (e.g. "Figure S1"). This is in case you want to add actual figures and not just captions.
\setcounter{figure}{0}
\renewcommand{\thefigure}{S\arabic{figure}}

%\clearpage

\section{Shape comparison procedure}
\label{sec:suppl_shape_comp}

\begin{figure}
\centering
\includegraphics[width=1.0\textwidth]{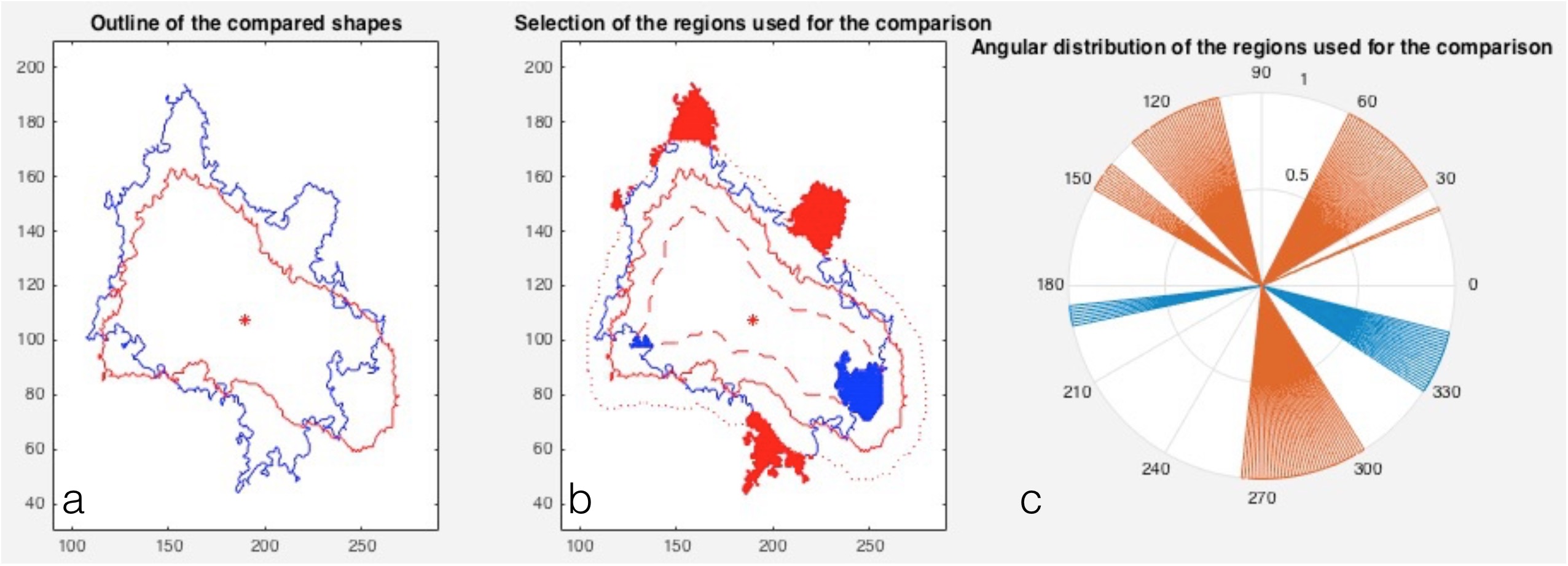}
\caption{Criterion developed to measure how similar two extinction spots are. The procedure is used to check whether two extinction spots, registered at the same sample rotation angle, belong to the same grain. ({\textbf{a}}) Perimeter of the two extinction spots A (in blue) and B (in red) (area(A)$>$area(B)), translated such that they have the same centre of mass. ({\textbf{b}}) Selection of the regions used to measure how similar the two shapes are: portions of eroded B outside A (in blue), and portions of A outside dilated B (in red). ({\textbf{c}}) Angular distribution of the located regions. Considering the angular range over which the two types of portions are distributed, the developed algorithm decides whether two extinction spots are similar or not.}
\label{sfig:shape_comparison}
\end{figure}

At a given projection, once the extinction spots are segmented it is necessary to find out which spots belong to the same grain. To do so, a criterion was developed to measure similarity between different spots. The approach, illustrated in Fig. \ref{sfig:shape_comparison}, employs the morphological operations of erosion and dilation \cite{russ2016image}. In a binary image, dilation adds pixels to the boundaries of shapes, while erosion removes pixels on shape boundaries. To compare two spots A and B (A being the one with the larger area), we superimposed their centers of mass and measured the angular distribution of the pixels of B that are inside the eroded perimeter of A, and of the pixels of A that are outside the dilated perimeter of B. If the angular regions where the spots differ are smaller than the chosen thresholds, the spots are considered similar and thus relative to the same grain.

\section{Curves fitting the extinction spots distribution}
\label{sec:suppl_calc_LO}

The crystallographic orientation of a given grain can be calculated (non uniquely, see Sec. \emph{U and CUC} below) by fitting the distribution of the relative combined extinction spots in the $\omega\lambda$-space, where $\omega$ is the sample rotation angle and $\lambda$ the wavelength relative to a combined extinction spot. As a starting point, let us consider Bragg's law 

\begin{equation}
\lambda = 2d\sin\theta
\end{equation}

and the diffraction equation in the form used by Poulsen et al. \cite{poulsen2001three}

\begin{equation}
  \textbf{G} = \frac{d}{2\pi}\Omega \text{\emph{U}}\mathcal{B}\textbf{h}
  \label{seq:fund_diff_eq}
\end{equation}

where $|\textbf{G}| = 1$, $d = \lambda/2\pi$ is the spacing between the lattice planes, $\Omega$ is the left-handed rotation matrix around the $z$-axis by an angle $\omega$, $\text{\emph{U}}$ is the orientation matrix and $\mathcal{B}$ is the matrix which maps the {\emph{hkl}} lattice, $\textbf{h} = \left(\begin{smallmatrix}h \\ k \\ l\end{smallmatrix}\right)$, into reciprocal space. Eq. (\ref{seq:fund_diff_eq}) can be rewritten as 

\begin{equation}
  \textbf{G} = \frac{\lambda}{4\pi}\Omega\text{\emph{U}}\mathcal{B}\textbf{h}
\label{seq:G_lambdaomega}
\end{equation}

Following the geometry sketched in Fig. \ref{fig:Detailed_geometry}, $\textbf{G}$ can be expressed \cite{schmidt2014grainspotter} as a function of the angles $2\theta$ and $\eta$

\begin{equation}
  \textbf{G} = \frac{1}{2} \left(
    \begin{array}{c}
      \cos2\theta - 1 \\
      -\sin2\theta\sin\eta \\
      \sin2\theta\sin\eta
    \end{array}
  \right)
  \label{seq:theta_eta}
\end{equation}

Considering the first component of $\textbf{G}$ and using Bragg's law, from Eq. (\ref{seq:theta_eta}) one has

\begin{equation}
\textbf{G}_1 = -\sin^2\theta = -\frac{\lambda^2}{4d^2}
\end{equation}

hence from Eq. (\ref{seq:G_lambdaomega})

\begin{equation}
  -\frac{\lambda^2}{4d^2} = \frac{\lambda}{4\pi} (\Omega \text{\emph{U}}\mathcal{B}\textbf{h})_1
\end{equation}

Writing $\lambda$ as a function of $\omega$ and introducing the vector $\textbf{v} = \mathcal{B}\textbf{h}$, one obtains the fundamental equation used to fit the point distribution in the $\omega\lambda$-space

\begin{align}
  \lambda(\omega) = - \frac{d^2}{\pi} (\Omega \text{\emph{U}}\mathcal{B}\textbf{h})_1 = - \frac{d^2}{\pi} (\Omega \text{\emph{U}}\textbf{v})_1
  \label{seq:omega_U_v}
\end{align}

Renormalizing the scattering vector to $|\textbf{G}| = \frac{\lambda}{2d}$ and considering the absolute value of Eq. (\ref{seq:G_lambdaomega}) leads to the expression $|\textbf{G}| = \frac{\lambda}{2d} = \frac{\lambda}{4\pi}|\mathcal{B}\textbf{h}|$.  Using this relation and developing all terms, Eq. (\ref{seq:omega_U_v}) becomes

\begin{equation}
  \begin{aligned}
    \lambda(\omega) = -\frac{4\pi}{|{\mathcal{B}\textbf{h}}|^2}\big[ (u_{11}v_1 + u_{12}v_2 + u_{13}v_3)\cos\omega +\\
    + (u_{21}v_1 + u_{22}v_2 + u_{23}v_3)\sin\omega \big]
  \end{aligned}
  \label{seq:start_forward}
\end{equation}

For a given reflection, corresponding to a given $hkl$ family, the coefficients $ A = u_{11}v_1 + u_{12}v_2 + u_{13}v_3$ and $B = u_{21}v_1 + u_{22}v_2 + u_{23}v_3$ have constant value, leading to the concise expression

\begin{equation}
\boxed{\lambda(\omega) = -\frac{4\pi}{|{\mathcal{B}\textbf{h}}|^2}(A \cos\omega + B \sin\omega)}
  \label{seq:final_OL}
\end{equation}

with the functions of $\omega$ being the only variables on the right-hand side.

\section{Grain indexing procedure}
\label{sec:suppl_index_proc}

\begin{figure}
\centering
\includegraphics[width=0.8\textwidth]{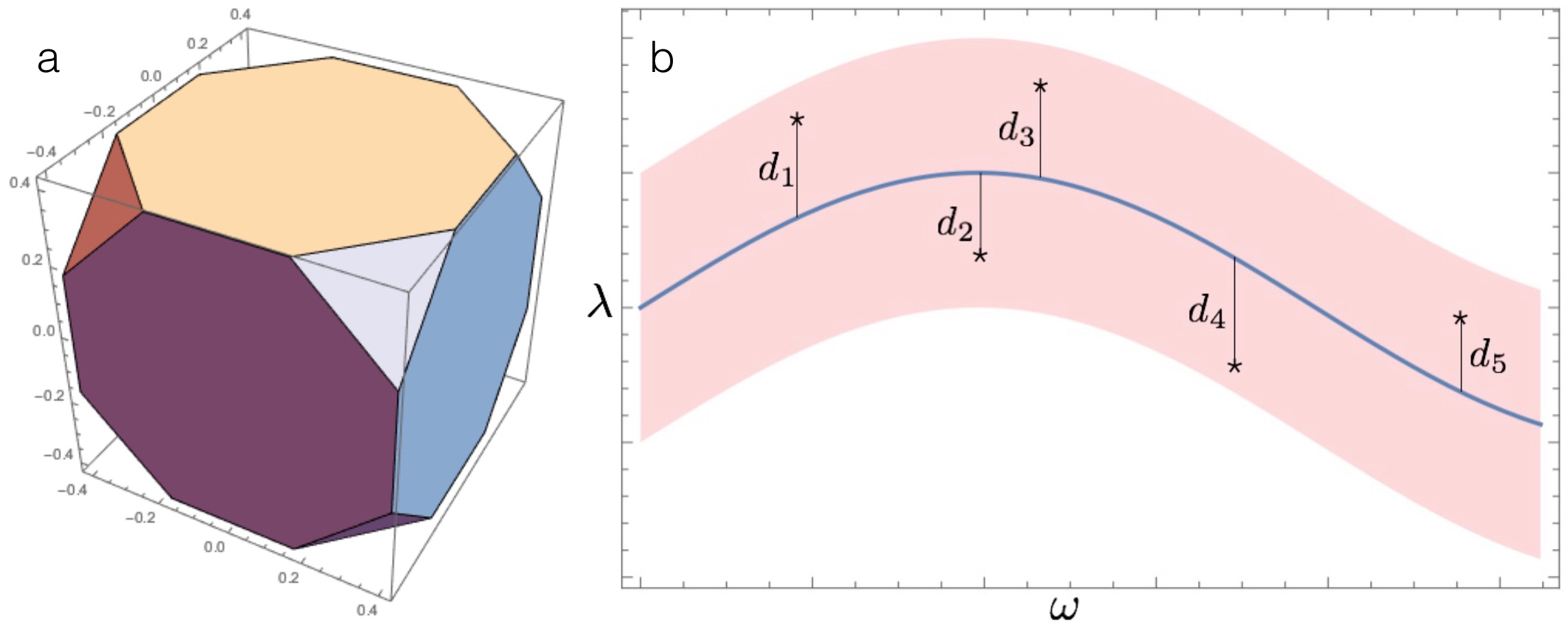}
\caption{Details of the indexing procedure. The orientation of the grains is calculated using a forward model, sampling a region of the Rodrigues space, where orientations are represented as points, and selecting the orientation best fitting the point distribution in the $\omega\lambda$-space (see Fig. \ref{fig:OL_raw_and_fit}), where $\omega$ is the sample rotation angle and $\lambda$ is the center of the wavelength interval where an extinction spot is recorded. ({\textbf{a}}) For a cubic system, all orientations can be considered by sampling the points of the {\emph{fundamental zone}}, a truncated cube with size $2\cdot(\sqrt{2}-1)$, whose corners are truncated by planes corresponding to rotations around (111) axes at a distance of $\tan(\pi/6)$ from the origin \cite{he2007representation, Kumar1995}. ({\textbf{b}}) To calculate the distance of a curve from the experimental values, the sum of the vertical distances $d_i$ is considered.}
\label{sfig:forward_model}
\end{figure}

This is the recipe followed to index (non uniquely, see Sec. \emph{U and CUC} below) a grain: 

\begin{enumerate}
\item Scan the Rodrigues space for possible orientations \cite{Morawiec_1996}. For a cubic system, all orientations can be considered by sampling the points of the {\emph{fundamental zone}}, a truncated cube with size $2\cdot(\sqrt{2}-1)$, whose corners are truncated by planes corresponding to rotations around (111) axes at a distance of $\tan(\pi/6)$ from the origin \cite{he2007representation, Kumar1995}. For each considered point (corresponding to an orientation), calculate the corresponding $\lambda(\omega)$ curves using Eq. (\ref{seq:final_OL}).
\item For each orientation, measure the distance of the fitting curves from the experimental values:
\begin{enumerate}
  \item For each curve, measure the vertical distance from it ($d_i$) of the points located in a band around it (see Fig. \ref{sfig:forward_model}).
  \item Considering all the different $\lambda(\omega)$ curves, calculate $D = \sum_i d_i$.
\end{enumerate}
\item Select the orientation whose $\lambda(\omega)$ curves return the minimum $D$.
\item Refine the selected orientation value by sampling, using a small grid, a region of the Rodrigues space built around the corresponding orientation vector.  
\end{enumerate}

For the Fe and the Co-Ni-Ga sample, grains are indexed considering the $hkl$ families 110, 200 and 211. The forward model is applied to values with $\lambda > 2$\AA: at lower wavelengths, the point distribution is more dense and harder to fit (see Fig. \ref{fig:OL_raw_and_fit}).

\section{U and CUC}
\label{sec:suppl_U_CUC}

For a given grain, fitting the distribution of its combined extinction spots using the forward model described in \ref{sec:suppl_index_proc} does not uniquely determine its orientation. The ambiguity is derived in the following. 

To be a proper rotation matrix, $U$ must satisfy the condition $\det(U) = 1$, with

\begin{equation}
  \begin{aligned}
    \det(U) = u_{11}(u_{22}u_{33} - u_{23}u_{32}) + u_{22}(u_{11}u_{33} - u_{13}u_{31}) + u_{33}(u_{11}u_{22} - u_{12}u_{21})
  \end{aligned}
  \label{seq:det_LOm}
\end{equation}

Considering the inversion $v_3 \rightarrow -v_3$, a change of sign of $v_3$ in Eq. (\ref{seq:start_forward}) results in the following changes:

\begin{enumerate}
  \item $u_{13}\rightarrow -u_{13}$ and $u_{23}\rightarrow -u_{23}$, to conserve Eq. (\ref{seq:start_forward}).
  \item $u_{31}\rightarrow -u_{31}$ and $u_{32}\rightarrow -u_{32}$, to conserve Eq. (\ref{seq:det_LOm}).
\end{enumerate}

In other words, changing the sign of $v_3$ results in the following changes of sign for the elements of the orientation matrix $U$

\begin{equation}
  U_{v_3} =
  \left(
  \begin{array}{ccc}
    + & + & + \\
    + & + & + \\
    + & + & +
  \end{array}
  \right) \rightarrow U_{-v_3} =
  \left(
  \begin{array}{ccc}
    + & + & - \\
    + & + & - \\
    - & - & +
  \end{array}
  \right)
\end{equation}

The two orientation matrices, $U_{v_3}$ and $U_{-v_3}$, and the relative Rodrigues vectors, $\textbf{r}_{v_3} = \left(\begin{smallmatrix}r_1 \\ r_2 \\ r_3 \end{smallmatrix}\right)$ and $\textbf{r}_{-v_3} = \left(\begin{smallmatrix} -r_1 \\ -r_2 \\ r_3 \end{smallmatrix}\right)$, are related by the expressions

\begin{eqnarray}
  U_{-v_3} & = & C U_{v_3} C^{-1}\\
  \textbf{r}_{-v_3} & = & C \textbf{r}_{v_3}
  \label{eq:C_r}
\end{eqnarray}

with

\begin{equation}
  C =
  \left(
  \begin{array}{ccc}
    -1 & 0 & 0 \\
    0 & -1 & 0 \\
    0 & 0 & 1
  \end{array}
  \right)
\end{equation}
\begin{equation}
  C = C^{-1}
\end{equation}

Consequently, based on transmission data alone the sign of the component of \textbf{G}, $v_3$, parallel to the z-rotation axis cannot be determined. In other words, flipping the direction of all the diffracted neutrons along the z-axis (from upwards to downwards and vice versa) will result in the same distribution in the $\omega\lambda$-plane. This ambiguity can be resolved by either locating a diffraction spot, whose position uniquely determines $v_3$, or by introducing a second rotation axis perpendicular to the z-axis as part of the data acquisition procedure. 

\section{Grain maps of the Co-Ni-Ga sample}
\label{sec:CoNiGa_grain_map}

\begin{figure}
\centering
\includegraphics[width=0.8\textwidth]{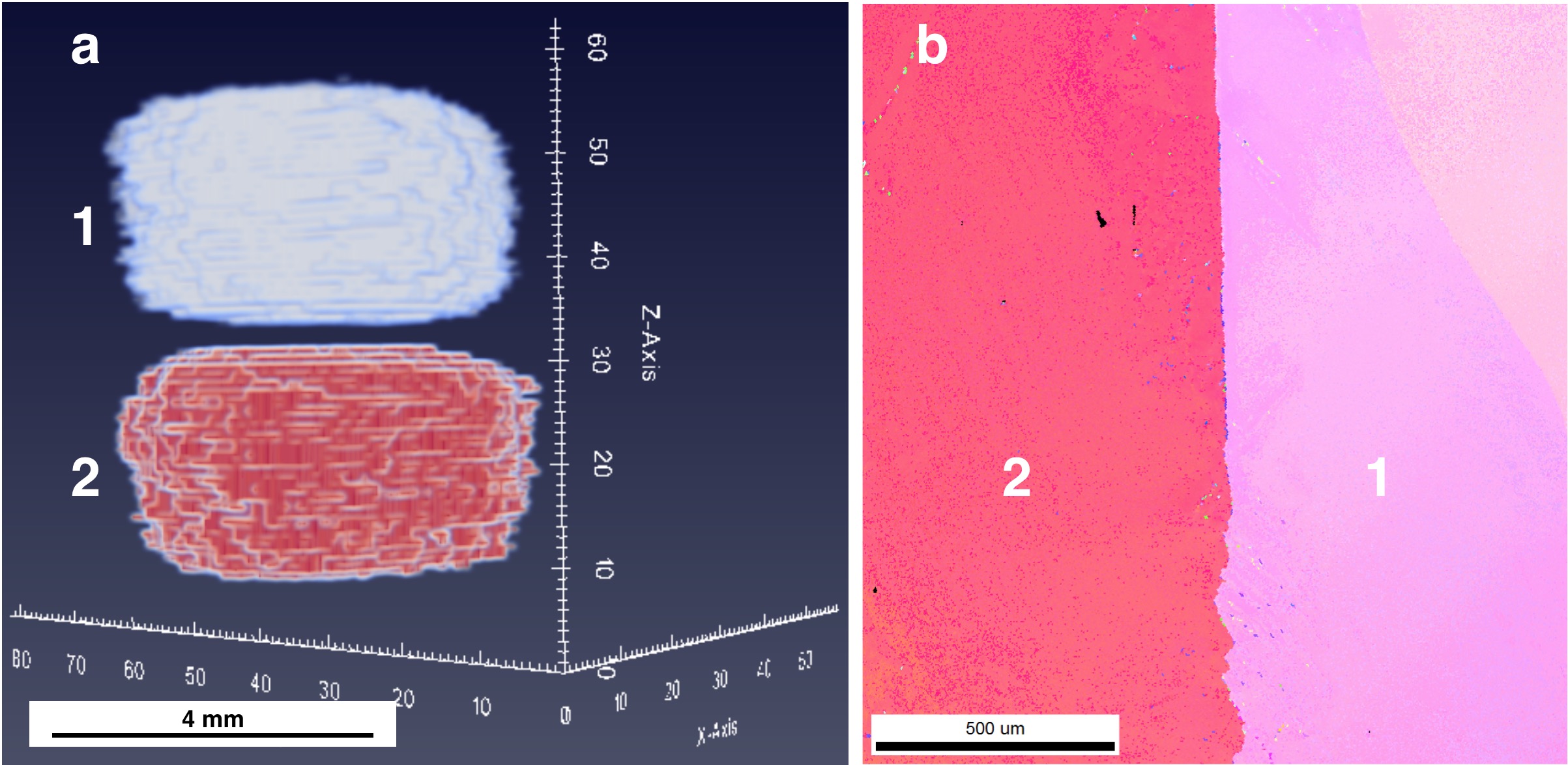}
\caption{Grain maps of the Co-Ni-Ga sample studied at {\footnotesize{SENJU}} to validate the {\footnotesize{ToF 3DND}} indexing procedure. ({\textbf{a}}) {\footnotesize{ToF 3DND}} reconstruction of the two grains (cubes with 4 mm side) in the Co-Ni-Ga sample. The spacing between grains is due to cutoffs when back-projecting the extinction spots. ({\textbf{b}}) {\footnotesize{EBSD}} map of a sample slice, modified from Vollmer {\emph{et al.}} \cite{vollmer2015damage} The grain numbering is the same in the two figures.}
\label{sfig:Bamboo_sample}
\end{figure}

The algorithms developed to reconstruct, from the transmission data, the {\footnotesize{3D}} shape of the grains and their orientation were validated by considering a Co-Ni-Ga sample, made of two single-crystal cubes, 4 mm in side. Fig. \ref{sfig:Bamboo_sample} shows two grain maps of the Co-Ni-Ga sample, obtained using {\footnotesize{ToF 3DND}} and {\footnotesize{EBSD}}.

%\nolinenumbers

%\bibliography{3DND}

%This defines the bibliographies style. Search online for a list of available styles.
\bibliographystyle{abbrv}

\end{document}